%% file: ProofStudio-ICSE-Demo.tex
\documentclass[10pt,conference]{IEEEtran}
\usepackage[T1]{fontenc}
\usepackage{cite}
\usepackage{balance}
\usepackage{amsmath,amssymb}
\usepackage{graphicx}
\usepackage{booktabs,tabularx,array}
\usepackage{listings}
\usepackage{xurl}
\usepackage{tikz}
\usetikzlibrary{arrows.meta,positioning,fit,calc}
\usepackage[hidelinks]{hyperref}
\usepackage{enumitem}
\input{submission-config}

\hypersetup{pdftitle={ProofStudio: Five Proof Assistants in the Browser},
 pdfauthor={Yutaka Nagashima},pdfsubject={ProofStudio; v2026-10-04.1}}
\newcommand{\tool}{ProofStudio}
\newcommand{\ui}[1]{\textsf{#1}}
\newcolumntype{Y}{>{\raggedright\arraybackslash}X}
\begin{document}
\title{ProofStudio: Five Proof Assistants in the Browser}
\author{\IEEEauthorblockN{Yutaka Nagashima}
\IEEEauthorblockA{Institute of Computer Science, The Czech Academy of Sciences}}
\maketitle
\begin{abstract}
Recent advances in AI-assisted proof automation are extending
the use of proof assistants in software verification and mathematics.
However, greater proof automation does not remove users' responsibility
to assess whether formal statements capture their intended
requirements, making education in formal specification and
reasoning essential.
To lower the installation and learning barriers associated with
mature proof assistants, we present ProofStudio, a browser-based
environment integrating Isabelle/HOL, Lean, Rocq, HOL4, and
HOL Light with corresponding worked examples, preconfigured
proof search, and contextual AI assistance.
With ProofStudio, users can explore formal specifications and proofs across
five systems without local installation, inspecting native prover
feedback and experimenting with automation through a common
interface.
The tool is available at \url{\ToolURL}.
\ifx\DemoVideoURL\empty
Demonstration video: \url{https://www.youtube.com/watch?v=AAj0tYQzZAA}.
\else
Demonstration video: \url{\DemoVideoURL}.
\fi
\end{abstract}

\begin{IEEEkeywords}
formal verification, proof assistants, web-based tools, proof automation
\end{IEEEkeywords}

\section{Introduction}\label{sec:intro}
Recent advances in AI-assisted proof automation are expanding
the practical scope of formal verification.
Anthropic and Theorem recently reported using Claude to produce
Rocq-checked proofs of no-crash properties for 239 entry points
across 123 open-source libraries, subject to explicit execution
assumptions and input bounds~\cite{anthropic-formal-verification}.
In mathematics, OpenAI released a large collection of
AI-generated manuscripts, many accompanied by Lean
formalizations~\cite{openai-math-release}.
These developments make proof assistants increasingly relevant
to software engineers and scientists beyond the formal methods
community.

Greater automation, however, does not remove the need to
understand and scrutinize what is being proved.
OpenAI withdrew three manuscripts after a sign error invalidated
an argument and two dependent proofs~\cite{openai-math-history}.
Beyond errors in reasoning, a separate concern remains even
when a formal proof is accepted: verification establishes that
a statement follows from its assumptions, whereas validation
asks whether that statement captures the intended requirement.
Anthropic's report explicitly identifies the adequacy of
specifications and machine models as limitations of its
guarantees~\cite{anthropic-formal-verification}.
Recent work on AI-generated mathematics likewise reports
discrepancies between natural-language arguments and their
formalizations~\cite{bastounis2026translation}.
Scientists and engineers therefore still need the skills to
read formal statements, inspect assumptions, and assess the
scope of automated results.

Developing these skills requires opportunities to experiment.
Mature proof assistants offer different logics, libraries,
proof languages, and automation, with complementary strengths.
However, trying each system often requires a separate installation
and editor configuration. These barriers affect newcomers and
make it harder even for experienced users to explore other provers.
Corresponding worked examples can help users compare how different
systems specify and verify familiar algorithms.

Local experimentation also requires trust in the installed
software.
Earlier Lean 4 installation instructions explicitly acknowledged
this requirement~\cite{lean4-install-historical}, while the
current guide includes a publisher-trust prompt for the Lean
extension~\cite{lean4-install-manual}.
Lean's official FAQ further warns that opening files, building
projects, or updating dependencies can execute arbitrary code,
and recommends isolation for potentially malicious
code~\cite{lean-security-faq}.
Remote hosting allows initial experimentation without installing
or executing prover software on users' own machines, shifting
responsibility for installation and execution containment
to the service.

We present \tool{}, a browser-based environment for exploring
Isabelle/HOL, Lean, Rocq, HOL4, and HOL Light.
Its contribution is the integration of:
\begin{itemize}[leftmargin=*,labelsep=0.5em]
    \item installation-free access to five native proof assistants
          through a common interface;
    \item ten corresponding examples per prover;
    \item preconfigured proof-search tools for supported provers; and
    \item contextual AI assistance with a choice of model provider.
\end{itemize}
Users edit native proof source, inspect intermediate proof states,
and review proposed source changes before accepting them.
The intended users are learners exploring formal reasoning and
experienced prover users trying another system.

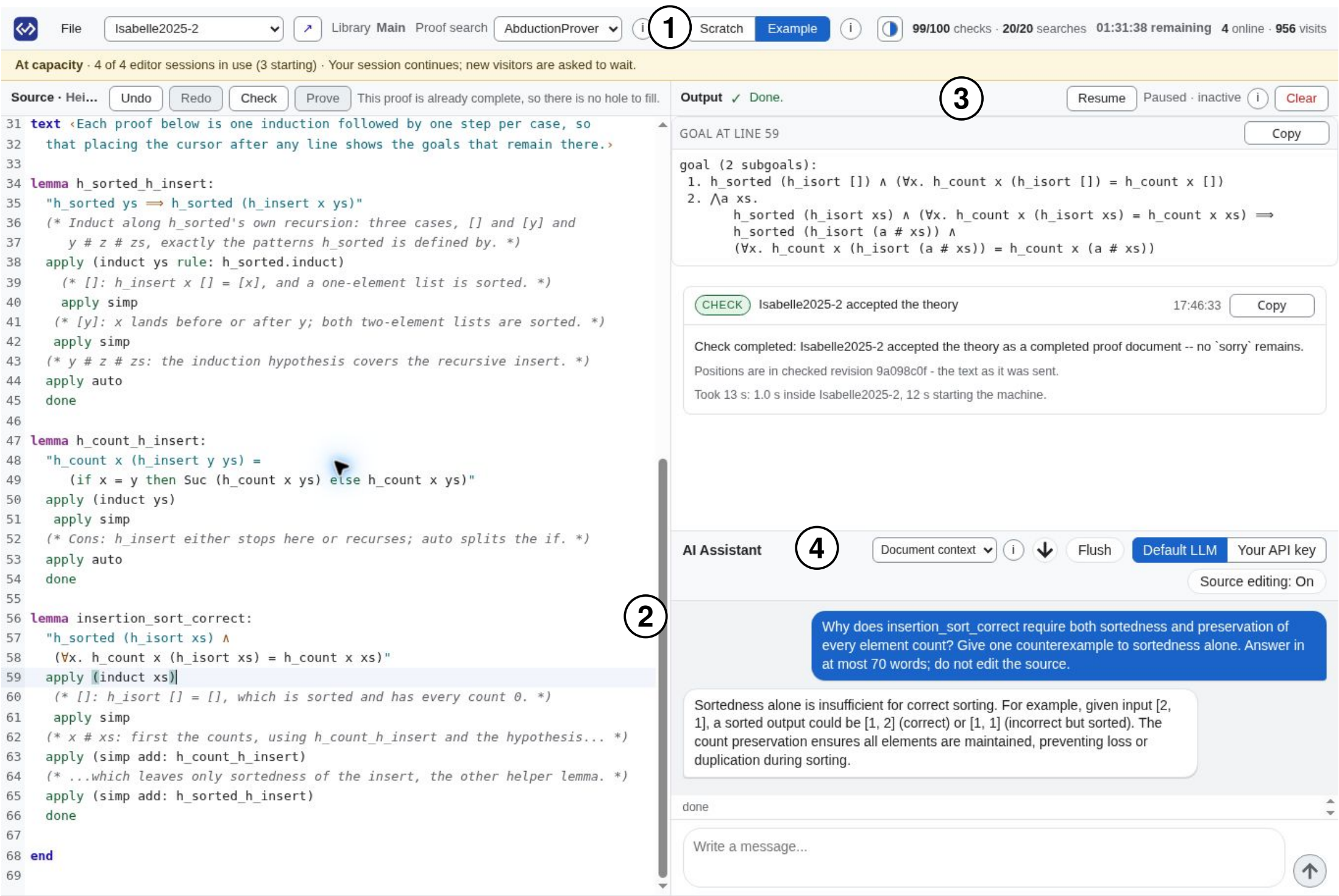
\begin{figure*}[t]
\centering
\begin{tikzpicture}
\node[anchor=south west,inner sep=0] (screen) at (0,0) {%
  \includegraphics[
    width=0.8\textwidth,
    trim=0 30 0 0,
    clip
  ]{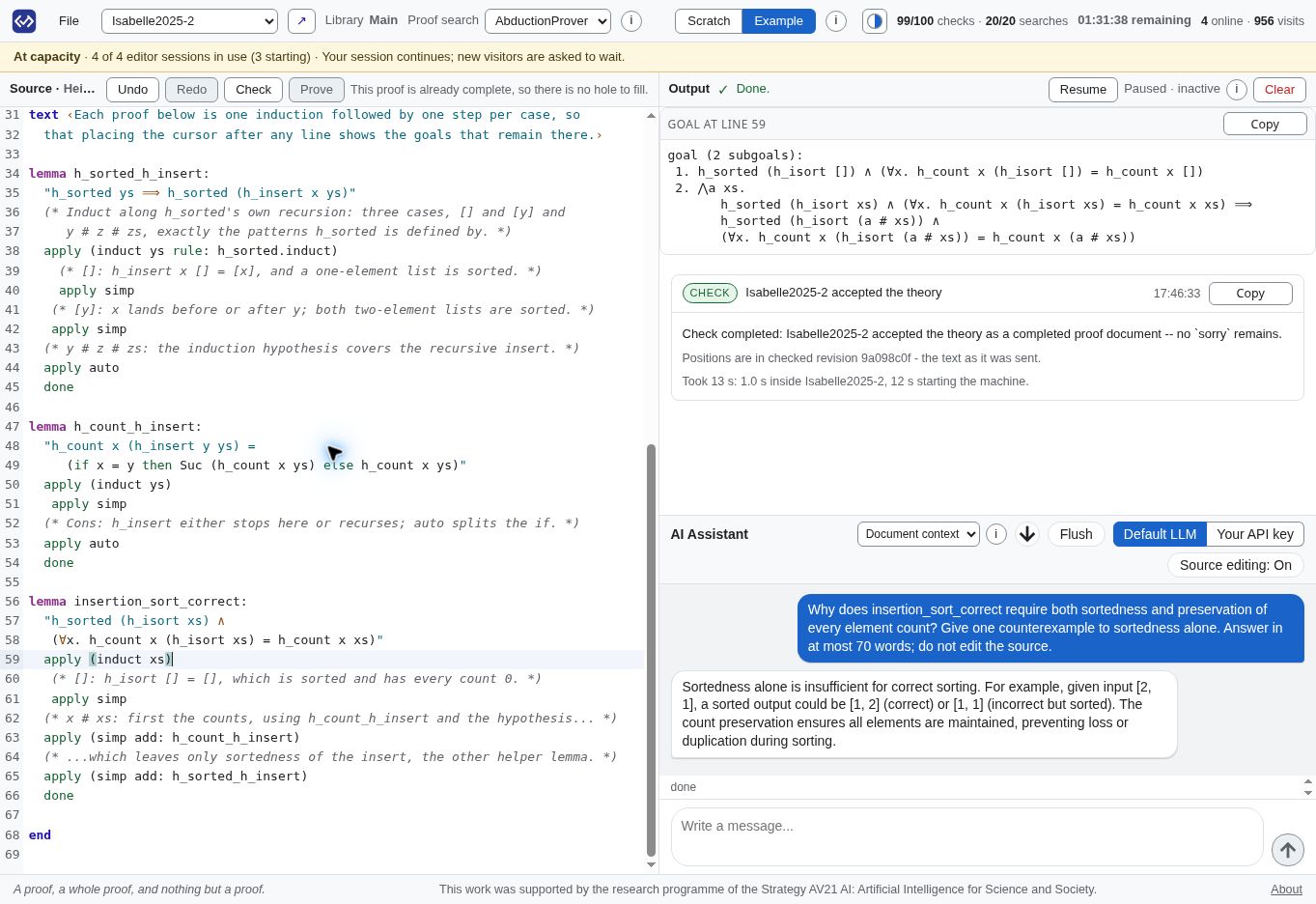}%
};

\begin{scope}[x={(screen.south east)},y={(screen.north west)}]
\tikzset{callout/.style={
  circle,
  draw=black,
  fill=white,
  line width=.7pt,
  font=\sffamily\bfseries\small,
  inner sep=1.7pt
}}
\node[callout] at (.500,.977) {1};
\node[callout] at (.482,.315) {2};
\node[callout] at (.718,.897) {3};
\node[callout] at (.610,.392) {4};
\end{scope}
\end{tikzpicture}

\caption{ProofStudio with the Isabelle insertion-sort example:
\textcircled{1} prover and example selection;
\textcircled{2} the \texttt{Source} pane;
\textcircled{3} the \texttt{Output} pane, showing intermediate subgoals and a successful
whole-document check;
\textcircled{4} the \texttt{AI Assistant} pane, explaining why correctness requires
both sortedness and count preservation.
Public-service capture, 21 October 2026.}
\label{fig:user-interface}
\end{figure*}

\section{User Experience}\label{sec:user-interface}
Fig.~\ref{fig:user-interface} shows the \texttt{Source}, \texttt{Output},
and \texttt{AI Assistant} panes. The menu selects the prover, example,
and search tool. At \url{\ToolURL}, information buttons beside
\texttt{Proof search} and \texttt{Output} provide usage guidance.

\subsection{Editing and Native Prover Feedback}
\label{sec:persistent}
The \texttt{Source} and \texttt{Output} panes support the basic cycle of
proof development: 
edit a definition or proof in the \texttt{Source} pane, 
then inspect the prover's response in the \texttt{Output} pane. 
The \texttt{Source} pane supports basic syntax highlighting and the
Unicode symbols used in the five proof assistants. 

Persistent prover sessions provide automatic feedback during editing, 
while the caret position identifies the part of the proof the user wants to inspect. 
In Fig.~\ref{fig:user-interface}, for example, 
the \texttt{Output} pane shows the two subgoals after applying the \texttt{induct} tactic, 
because the caret is positioned immediately after induction on \texttt{xs}. 
These describe the local proof state at that point.
The subsequent commands shown in Fig.~\ref{fig:user-interface}
complete the proof.

\subsection{Checking Documents and Searching for Proofs}
\label{sec:contracts}
The \texttt{Source} pane offers two explicit commands: \texttt{Check} and \texttt{Prove}.
\texttt{Check} requests native diagnostics and proof information, including
feedback on unfinished documents. 
Successful processing is distinct from proof completion:
if remaining goals are admitted using commands such as 
\texttt{sorry} in Lean and Isabelle or \texttt{Admitted} in Rocq,
\texttt{Check} accepts the document even when proofs are still incomplete.

When the \texttt{Source} pane contains such holes,
users can try the \texttt{Prove} button,
which searches for a proof script that fills a proof hole.
Table~\ref{tab:capabilities} summarizes the available search options for \texttt{Prove},
which users can choose in the main menu bar at the top.
These include AbductionProver~\cite{abductionprover} for Isabelle,
Aesop~\cite{aesop} for Lean, CoqHammer~\cite{coqhammer} for Rocq, and
TacticToe~\cite{tactictoe} for HOL4. 
The current \texttt{Prove} command uses conventional proof search;
LLM-based assistance is provided separately through the assistant pane.

Upon successful proof search, 
the \texttt{Prove} command checks that source outside the hole is unchanged 
and uses a separate verifier to replay and audit the target theorem,
as shown in Fig.~\ref{fig:architecture} in Section~\ref{sec:architecture}.

\begin{table}[t]
\caption{Default offering in \ReleaseTag{}. All five support Check.
Library scope excludes separately bundled search tools.}
\label{tab:capabilities}
\centering\footnotesize
\begin{tabularx}{\columnwidth}{@{}l l Y@{}}
\toprule
Prover & Library scope & Prove options \\
\midrule
Isabelle/HOL & Main & AbductionProver, Sledgehammer, \texttt{try0} \\
Lean 4 & Core & Portfolio, Aesop, \texttt{grind} \\
Rocq & Stdlib & CoqHammer \\
HOL4 & Core & TacticToe, standard automation \\
HOL Light & Core & Not supported \\
\bottomrule
\end{tabularx}
\end{table}

\subsection{Corresponding Examples Across Five Provers}

\tool{} provides ten corresponding examples for each prover:
insertion sort, merging, ordered-list sets, binary search trees,
two-list queues, binary tries, leftist heaps, binomial-tree linking,
Braun trees, and Fibonacci pairs. Adapted from
\emph{Functional Data Structures and Algorithms: A Proof Assistant
Approach}~\cite{fdsa}, these self-contained developments let users
compare native specifications and proof techniques without configuring
additional libraries. 
For the insertion-sort example in Fig.~\ref{fig:user-interface},
users can keep sortedness and preservation of element counts
as the intended requirements while exploring how each prover
expresses and establishes them.

\subsection{AI Assistance and Provider Choice}
\label{sec:assistant}

All five provers share an AI Assistant interface with two modes: 
the \texttt{document context} mode and \texttt{question only} mode. 
In \texttt{document context} mode, the manager combines the current source, available goals, caret position, output, and conversation history with prover-specific instructions. 
This is the mode shown in Fig.~\ref{fig:user-interface}, where the assistant explains the two
parts of the sorting contract. 
In \texttt{question only} mode, it omits document context and disables AI-generated source edits.

Users can choose the shared default LLM or an external service
through bring-your-own-key (BYOK) configuration. BYOK requires
an API key and a public HTTPS endpoint supporting the
OpenAI-compatible interface. Section~\ref{sec:credentials}
explains the associated credential handling.

The model has no direct access to shell or prover tools. Proposed source
edits are held for user review, and accepting a proposal does not certify
its correctness. Users must inspect any changed specification and check the
result with the prover.

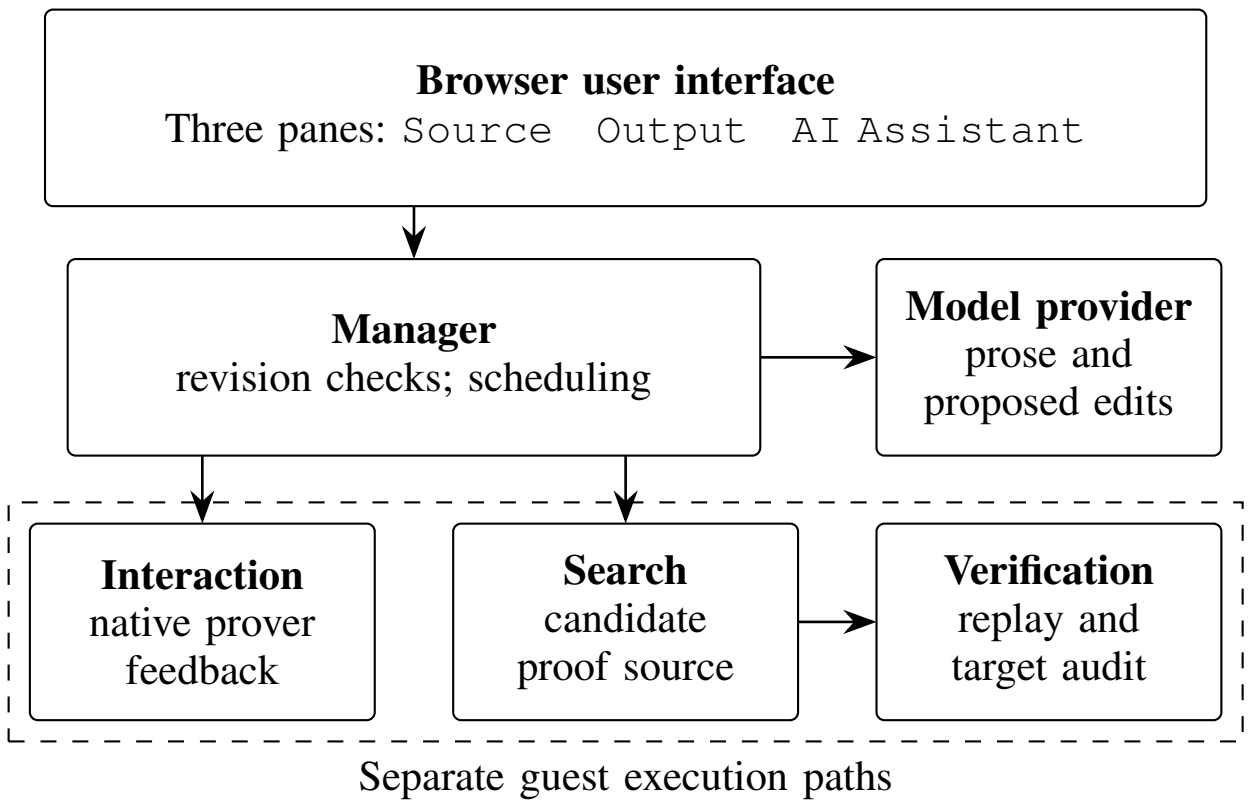
\begin{figure}[t]
\centering
\begin{tikzpicture}[font=\footnotesize,
 box/.style={draw,rounded corners=1.5pt,align=center,inner sep=4pt,
             outer sep=0pt,minimum height=13mm},
 arr/.style={-{Stealth[length=2.2mm,width=1.6mm]},line width=.5pt}]
\node[box,text width=74mm] (ui) at (0,0)
 {\textbf{Browser user interface}\\ Three panes: \texttt{Source}\quad\texttt{Output}\quad\texttt{AI Assistant}};
\node[box,text width=43mm] (manager) at (-1.4,-1.65)
 {\textbf{Manager}\\revision checks; scheduling};
\node[box,text width=20mm] (model) at (2.8,-1.65)
 {\textbf{Model provider}\\prose and proposed edits};
\node[box,text width=20mm] (check) at (-2.8,-3.4)
 {\textbf{Interaction}\\native prover\\feedback};
\node[box,text width=20mm] (search) at (0,-3.4)
 {\textbf{Search}\\candidate\\proof source};
\node[box,text width=20mm] (verify) at (2.8,-3.4)
 {\textbf{Verification}\\replay and\\target audit};
\draw[arr] (ui.south -| manager.north) -- (manager.north);
\draw[arr] (manager.east) -- (model.west);
\draw[arr] (manager.south -| check.north) -- (check.north);
\draw[arr] (manager.south -| search.north) -- (search.north);
\draw[arr] (search.east) -- (verify.west);
\node[draw,dashed,fit=(check)(search)(verify),inner sep=4pt,
 label={[font=\footnotesize]below:Separate guest execution paths}] {};
\end{tikzpicture}
\caption{Logical execution paths; workers and return paths are omitted.
Results return through the manager. AI edits require user review;
repairs produced by \texttt{Prove} undergo source-preservation checks
and prover-specific verification. Guest containment has the limits discussed
in Section~\ref{sec:isolation}.}
\label{fig:architecture}
\end{figure}

\section{System Architecture}
\label{sec:architecture}

Fig.~\ref{fig:architecture} presents the architecture of ProofStudio.
The manager coordinates native prover interaction, automatic proof
search, independent checking of search results, and access to
AI model providers.

\subsection{Prover Containment and Limitations}\label{sec:isolation}
Processing user-supplied proof documents can execute potentially
malicious code. Lean's FAQ warns about this risk~\cite{lean-security-faq}.
Remote hosting therefore requires containment of potentially
hostile computations.

ProofStudio uses QEMU/KVM virtual machines, which provide
hardware-assisted virtualization on Linux.
Each proof assistant instance runs inside a virtual machine,
while the manager serves the browser interface and schedules
workers that control prover execution.

Prover VMs have no virtual network interfaces.
During interactive proof development, workers exchange prover input
and output through a dedicated communication channel provided by QEMU.
Users interact with the provers through ProofStudio's web interface.
ProofStudio also screens submitted source to restrict accepted
commands and imports, limiting access to functionality beyond
the intended scope of introductory proof development.

These measures reduce the exposed functionality but do not establish
formally verified isolation.
PwnEscape (CVE-2026-99999) illustrates the dependence of VM isolation
on surrounding host software: a Linux networking flaw enabled
reported guest-to-host escape on the researchers' test
systems~\cite{pwnescape}.
The reported virtual-network attack path is absent from our
configuration, but other host-guest interfaces remain part
of the trusted execution boundary.

\subsection{Model Serving and Credential Handling}
\label{sec:credentials}
The default deployment serves Qwen3-Coder-30B-A3B-Instruct~\cite{qwen3coder}
through \texttt{llama.cpp} on a separate inference server.
This backend reflects the current resource budget and can be
replaced without changing the prover integrations.

Requests to both the default and external providers pass through
the manager, which provides common handling of document context,
streamed responses, and proposed source edits
(Fig.~\ref{fig:architecture}).
This centralized design also requires the manager to handle
the API keys supplied by BYOK users.

The security breach at OpenCode Console in October 2026
highlights the risks of handling users' API
keys~\cite{opencode-console-breach}.
To limit credential exposure, ProofStudio retains keys only
temporarily in server memory, without persistent storage,
and ends retained access when the associated session expires.
The default session lasts 100 minutes; this design still requires
users to trust the service while their keys are in use.
Its privacy policy, which users must accept before submitting
proof documents, also recommends disabling or rotating API keys
with their provider after use~\cite{proofstudio-privacy}.

\subsection{Standby Allocation Under Resource Constraints}
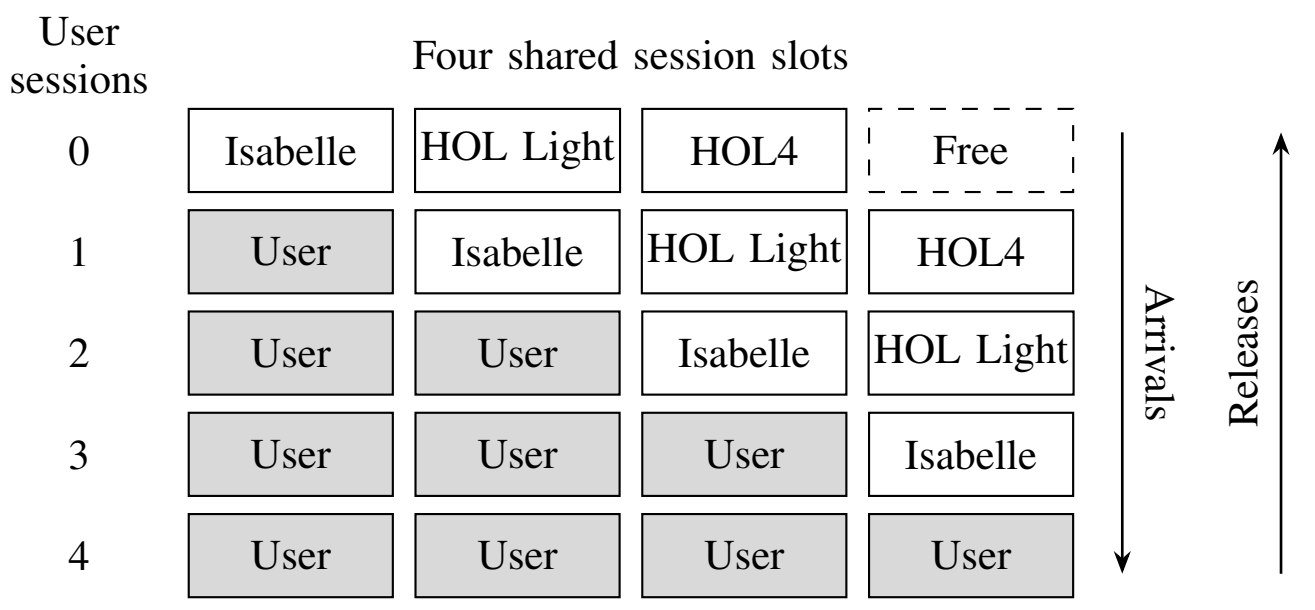
\begin{figure}[t]
\centering
\begin{tikzpicture}[font=\footnotesize,
 cell/.style={draw,minimum width=13.5mm,minimum height=5.5mm,
              inner sep=1pt,align=center},
 user/.style={cell,fill=black!15},
 spare/.style={cell},
 free/.style={cell,dashed},
 arr/.style={-{Stealth[length=1.5mm]},line width=.5pt}]
\node[align=center] at (0,.63) {User\\sessions};
\node at (3.65,.63) {Four shared session slots};
\foreach \n/\y in {0/0,1/-.67,2/-1.34,3/-2.01,4/-2.68}
  \node at (0,\y) {\n};
\node[spare] at (1.4,0) {Isabelle};
\node[spare] at (2.9,0) {HOL Light};
\node[spare] at (4.4,0) {HOL4};
\node[free] at (5.9,0) {Free};
\node[user] at (1.4,-.67) {User};
\node[spare] at (2.9,-.67) {Isabelle};
\node[spare] at (4.4,-.67) {HOL Light};
\node[spare] at (5.9,-.67) {HOL4};
\node[user] at (1.4,-1.34) {User};
\node[user] at (2.9,-1.34) {User};
\node[spare] at (4.4,-1.34) {Isabelle};
\node[spare] at (5.9,-1.34) {HOL Light};
\node[user] at (1.4,-2.01) {User};
\node[user] at (2.9,-2.01) {User};
\node[user] at (4.4,-2.01) {User};
\node[spare] at (5.9,-2.01) {Isabelle};
\foreach \x in {1.4,2.9,4.4,5.9}
  \node[user] at (\x,-2.68) {User};
\draw[arr] (6.9,.12) -- (6.9,-2.8);
\node[rotate=270] at (7.15,-1.34) {Arrivals};
\draw[arr] (7.95,-2.8) -- (7.95,.12);
\node[rotate=90] at (7.7,-1.34) {Releases};
\end{tikzpicture}
\caption{Desired allocation under the current resource-constrained default.
White named cells are standbys; shaded cells are user sessions of any
prover. Rows show targets after replenishment, not instantaneous readiness
or persistent slot identities. Booting and teardown also occupy capacity.}
\label{fig:standbys}
\end{figure}

To reduce startup delays, the default configuration for a server
with 64\,GiB of memory shares four configurable slots between user
sessions and standbys. As Fig.~\ref{fig:standbys} shows, with
$n\in\{0,\ldots,4\}$ allocated sessions, including those connecting,
the target is $\min(3,4-n)$ standbys, prioritized as Isabelle,
HOL Light, then HOL4. Isabelle and HOL Light initialize both the VM
and prover; HOL4 initializes only the VM. Lean and Rocq start on
demand because their prover initialization is comparatively inexpensive.

An arrival takes a matching standby, waiting if it is still booting,
or starts a new VM. Only unused standbys may be reclaimed;
assigned sessions are protected. Released VMs are destroyed and
replacements start from a clean state.
Replenishment requires memory headroom, boots one standby at a time,
and waits for a default 30\,s interval without new arrivals or
\ui{Check}/\ui{Prove} jobs. Standbys serve interactive sessions only;
these explicit jobs have no separate standby pools by default.
The policy reduces startup work but cannot guarantee a warm instance
for every arrival.

\section{Discussion and Conclusion}
\label{sec:conclusion}
\noindent\textit{Related work.}
Generic proof interfaces predate \tool{}: Proof General provides a common
interface to diverse proof assistants~\cite{proofgeneral}.
Browser interfaces also have an established history.
ProofWeb supports logic teaching through Coq~\cite{proofweb};
jsCoq runs Coq in the browser~\cite{jscoq};
Clide connects web-based editing to Isabelle's asynchronous
document model~\cite{clide}.
Lean4web offers a server-backed Lean editor for small developments
and demonstrations~\cite{lean4web}.
\tool{} combines five native prover environments, corresponding examples, preconfigured automation, and
AI assistance in one service.

\noindent\textit{Early use and proposed evaluation.}
A doctoral student with a mathematics background is exploring
\tool{} to learn Lean. This informal use has not produced systematic
evaluation results. Classroom use is under discussion; we welcome
collaboration with educators and researchers. A proposed study would
ask learners to inspect an example's specification, explain an
intermediate goal, and complete a proof hole, then compare a
corresponding example in another prover. Task completion, time,
errors, and short explanations of the specification would inform
usability and comprehension assessment. A classroom deployment
should use a dedicated instance provisioned for its expected
concurrency. 

\noindent\textit{Limitations and future work.}
The current scope is single-document developments with the library
coverage in Table~\ref{tab:capabilities}. Public-service capacity is
limited. Future work includes multi-document projects, broader
libraries, and increased capacity. 
Further development and evaluation would benefit from research
collaboration and additional computational resources.

\noindent\textit{Conclusion.}
\tool{} makes five proof assistants accessible through a common
browser workflow.
Users can explore the same algorithmic question across provers,
inspect intermediate proof states, and review AI suggestions
against native prover feedback.
The service complements dedicated prover IDEs by lowering the
barrier to exploration while keeping specifications and proof
evidence visible.

\section*{Acknowledgment.}
We used ChatGPT, Claude, Codex, Claude Code, OpenCode, and Kimi-K3 to assist with drafting and revising passages throughout the manuscript, implementing ProofStudio, and preparing and producing the demonstration video. We used Lean, HOL Light, HOL4, Rocq, and Isabelle/HOL for proof development and demonstrations. The author takes responsibility for the final content.
This work was supported by the research programme of the Strategy AV21 AI: Artificial Intelligence for Science and Society,
with institutional support RVO:67985807.

\section*{Data Availability}
The public service, in-interface usage guidance, and ten example
proof scripts for each of the five provers are available at
\url{\ToolURL}. 
Implementation source is available to reviewers upon request.
We intend to release the entire software stack as open source.
However, recent security incidents, including the OpenCode Console
breach~\cite{opencode-console-breach}, have heightened our concern
about protecting users' API credentials. We are therefore postponing
public source release until after internal penetration testing.

\bibliographystyle{IEEEtran}
\bibliography{references}
\end{document}

%% file: submission-config.tex
\def\DemoVideoURL{https://www.youtube.com/watch?v=AAj0tYQzZAA}
\ifx\DemoVideoURL\empty
  \PackageWarningNoLine{proofstudio}{DemoVideoURL is empty; draft placeholder will be printed}
\fi
\newcommand{\ToolURL}{https://proofstudio.org/}

\newcommand{\ReleaseTag}{v2026-10-04.1}